\documentclass[a4paper,twocolumn,showpacs,amsmath,prl,amssymb]{revtex4-1} %letterpaper,preprint
\usepackage[T1]{fontenc}
\usepackage[latin9]{inputenc}
\usepackage{times}
\usepackage{color}
\usepackage{xspace}
\usepackage{amssymb,amsmath}
\usepackage{amsbsy}
\usepackage{graphicx}
\usepackage{epstopdf}
\usepackage{float}

\newcommand{\diff}{\mathrm{d}}

\newcommand{\bx}{\ensuremath{\mathbf{x}}\xspace}
\newcommand{\x}{\mathbf{x}}

\newcommand{\Cdd}{C_{\textrm{dd}}}
\newcommand{\Udd}{U_{\textrm{dd}}}
\newcommand{\Utdd}{\tilde{U}_{\textrm{dd}}}
\newcommand{\xd}{\mathbf{x}^\prime}

\newcommand{\ldb}{\lambda_{\rm{dB}}\xspace}
\begin{document}
\newcommand{\bk}{\mathbf{k}}
\newcommand{\nt}{\tilde{n}}
\newcommand{\Lnh}{\ensuremath{\mathcal{L}}\xspace}
\newcommand{\dbx}{\diff\bx}
\newcommand{\dbk}{\diff\bk}
\newcommand{\nbe}{\ensuremath{\bar{n}_{\mathrm{BE}}}\xspace}

\title{Finite temperature stability of a trapped dipolar Bose gas}
\author{R.~N.~Bisset} 
\author{D.~Baillie} 
\author{P.~B.~Blakie}  

\affiliation{Jack Dodd Centre for Quantum Technology, Department of Physics, University of Otago, Dunedin, New Zealand.}

%\affiliation{Jack Dodd Centre for Quantum Technology, Department of Physics, University of Otago, Dunedin, New Zealand.}

\begin{abstract}
 We calculate the stability diagram for a  trapped normal Bose gas with dipole-dipole interactions. Our study characterizes the roles of  trap geometry, temperature, and short-ranged interactions on the stability. We predict a robust double instability feature in oblate trapping geometries arising from the interplay of thermal gas saturation and the anisotropy of the interaction. Our results are relevant to current experiments with polar molecules and will be useful in developing strategies to obtain a polar molecule Bose-Einstein condensate.

\end{abstract}
%\pacs{03.75.Ss, 74.20.Rp, 67.30.H-,05.30.Fk}

\maketitle

%==============================================================================
{\bf Introduction:}  
There has been phenomenal recent progress in experimental efforts to produce quantum degenerate  gases with   dipole-dipole interactions (DDIs) \cite{Stuhler2005a,Lahaye2007a,Ni2008a}. The interest in these systems is being driven by a broad range of proposed applications from condensed matter physics to quantum information, e.g.~see 
\cite{Kawaguchi2006a,Baranov2002a,*Goral2002a,*DeMille2002a,*ODell2003a,*Rabl2006a,*Buchler2007a}.
For bosonic dipolar gases the majority of theoretical work has been undertaken for  the $T=0$ case described by the Gross-Pitaevskii equation (e.g.~see \cite{Kawaguchi2006a,Goral2000a,*Kawaguchi2006b,*Ronen2006b,*Kawaguchi2007a,*Wilson2009a,*Parker2009a}), motivated by beautiful experiments with $^{52}$Cr  condensates \cite{Griesmaier2005a,*Bismut2010a,*Pasquiou2011a}. 
On the horizon is the realization of samples of degenerate polar molecules which have DDIs several orders of magnitude larger than those of the atomic systems. 
Currently bosonic \cite{Aikawa2010a} and fermionic \cite{Ni2008a} KRb molecules have been produced in their rovibrational ground state  and  effort is now focused on cooling these to  quantum degeneracy.

In addition to being a long ranged interaction, the DDI is also anisotropic with an attractive component. Thus an important consideration is under what conditions the system is mechanically stable from collapse to a high density state. Theoretical studies have been performed \cite{Ronen2007a} for the ($T=0$) condensate, in which this stabilization arises from the \textit{quantum pressure} of confinement (also from repulsive short range interactions \cite{Lu2010a}). Good agreement has been obtained between such theories and experiments with $^{52}$Cr \cite{Koch2008a}. 
 The stability of dipolar condensates is considerably richer than that predicted for a gas with attractive contact interactions. Notably,  
due to the DDI anisotropy stability  depends strongly on the geometry of the trapping potential \cite{Yi2001a,Eberlein2005a}. We note that detailed studies reveal rich nonlinear behavior (e.g.~spatially oscillating condensate density)  in parameter regimes near the boundary of instability related to the emergence of roton-like excitations \cite{Ronen2007a,Ronen2007b,Lu2010a}. An experimental study of condensate collapse dynamics has also been reported \cite{Lahaye2009a}.

Stability at finite temperature, pertinent to current experimental work with polar molecules, remains much less clear. In particular, while there has been some work on stability of a normal dipolar Fermi gas \cite{Zhang2010a}, the finite temperature bosonic system remains largely unexplored. In this paper we develop a theory for the  stability of a trapped dipolar Bose gas at temperatures above the critical temperature. Our work is based on self-consistent  meanfield calculations in which we identify the stability regime using the density response function, closely related to the system compressibility.  This  allows us to quantify the roles of trap geometry, temperature, and short range  interactions. %on the stability of this system. 

{\bf Formalism:} 
We consider a gas of bosons that interact by both a long-range DDI and a contact interaction characterized by  $U_{\rm{int}}(\x)=g\delta(\x)+U_{\rm{dd}}(\x)$ \cite{Yi2001a}, where $g=4\pi a\hbar^2/m$, with $a$ the s-wave scattering length.
We take the dipoles to be polarized along $z$ giving a dipole interaction of
\begin{equation}
U_{\rm{dd}}(\x)=\frac{C_{\rm{dd}}}{4\pi}\frac{1-3\cos^2\theta}{|\x|^3},\label{eqnUdd}
\end{equation}
with $\theta$ the angle between  $\x$  and the $z$ axis.  
%The constant $C_{\rm{dd}}$, given by $d^2/\epsilon_0$ ($\mu_0\mu^2_m$) for electric (magnetic) dipoles of strength $d$ ($\mu_m$).
The constant $C_{\rm{dd}}$, given by $\mu_0\mu^2_m$ for magnetic dipoles of strength $\mu_m$ and $d^2/\epsilon_0$ for electric dipoles of strength $d$.
The atoms are confined within a cylindrically symmetric harmonic trap
$U(\x)=\frac{m}{2}\left[\omega_\rho^2(x^2+y^2)+\omega_z^2z^2\right],$ with aspect ratio $\lambda=\omega_z/\omega_\rho$.
%The constant $C_{\rm{dd}}$, given by $\mu_0\mu^2_m$ for magnetic dipoles of strength $\mu_m$ and $d^2/\epsilon_0$ for electric dipoles of strength $d$.
 
Our primary concern is the properties of this system above the critical temperature for which we use   self-consistent meanfield theory \footnote{An  overview of the applicability of such theory in the regime we consider is given in Ref.~\cite{Kestner2010a}.}  giving the density as 
\begin{equation}
n(\x)=\ldb^{-3}\zeta_{3/2}^+\left(e^{\beta[\mu-V_{\rm{eff}}(\x)]}\right),\label{den}
\end{equation}
where $\mu$ is the chemical potential, $\beta=1/k_BT$ is the inverse temperature, $\zeta_{\alpha}^{\eta}(z)=\sum_{j=1}^{\infty}\eta^{j-1}z^j/j^\alpha$ is the Bose/Fermi function, and $\ldb=h/\sqrt{2\pi mk_BT}$.
The effective potential,
\begin{equation}
V_{\rm{eff}}(\x)=U(\x)+2gn(\x)+\Phi_D(\x),
\end{equation}
includes direct and exchange terms for the contact interaction, but only the direct term for the dipole interaction, given by
\begin{align}
\Phi_D(\x)&=\int  {d\xd  }\, U_{\rm{dd}}(\x-\xd)n(\xd). \label{PhiD}
\end{align}
The neglect of dipole exchange is consistent with other work on finite temperature bosons \cite{Ronen2007b} and zero temperature studies of fermion stability \cite{He2008a}.
Dipolar exchange has recently been included in equilibrium calculations for the fermionic system \cite{Zhang2010a} and found to be less significant than direct interactions except in near-spherical traps \cite{Baillie2010b}. More generally, the experience from  fermion studies suggests that exchange effects will give rise to shifts in the stability boundaries, but not change the overall qualitative behavior \cite{He2008a,Zhang2010a}.  
We also neglect collisional loss, such as exothermic  bimolecular reactions. This is an issue for reactive molecules such as KRb \cite{Ospelkaus2010a} but is expected to be unimportant for other alkali-metal dimers \cite{Zuchowski2010a} being pursued in experiments, such as RbCs \cite{Hudson2008a,*Lercher2011a}. 

The cylindrical symmetry of the system allows an accurate numerical treatment using Fourier-Bessel techniques \cite{Ronen2006a}. The main challenge in solving the mean field theory is that the numerical grids must have sufficient range and point density to accurately represent the DDI. In practice the grids become quite large for anisotropic traps. Another requirement is that Eqs.~(\ref{den})-(\ref{PhiD}) need to be solved self-consistently, which we implement using a fixed point iteration scheme. Near instability regions, the convergence rate of this  scheme generally decreases significantly.

 To calculate the stability phase diagram we determine the parameter regime where self-consistent meanfield solutions are obtained.
In practice we see a number of signatures of instability of our solutions, such as the divergence of the density (i.e.~density spike) and strong dependence on the numerical grid.
While such failures of convergence have been widely used to identify meanfield instability, for the results we present here we use an unambiguous condition in terms 
of  the density response function diverging   \footnote{We have numerically confirmed that convergence failure and the response function divergence are in excellent agreement for a range of systems with $0.1\lesssim\lambda\lesssim4$. Verifying agreement for large $\lambda$ becomes computationally demanding. }. This divergence is related to the mode-softening used in calculations by Ronen \textit{et al.}~to identify dipolar condensate instability \cite{Ronen2007a}.

The density response function of the system, in the random phase approximation (RPA), is given by
\begin{equation}
\chi(\mathbf{k})=\frac{\chi_0(\mathbf{k})}{1+[2g+\Utdd(\mathbf{k})]\chi_0(\mathbf{k})}.\label{chi}
\end{equation}
Here $\chi_0(\mathbf{k})$ is the bare response function and $\Utdd(\mathbf{k}) = \Cdd(\cos^2\theta_k-1/3)$ is the Fourier transform of $\Udd(\x)$. Within the semiclassical approximation used in our meanfield theory the response functions also depend on position, 
however as  instability occurs at trap center we take these to be evaluated at $\x=\mathbf{0}$.
We note that Eq.~(\ref{chi}) includes direct and exchange contact interactions \cite{Minguzzi1997a}, but only the direct dipolar interaction (also see \cite{Yamaguchi2010a}).
For  stability considerations we take $\bk\to\mathbf{0}$ along the direction which $\Utdd(\bk)$ is most attractive, i.e.~$\theta_k=\pi/2$. With this limit the stability region is determined by the condition 
\begin{equation}
1+\left(2g- C_{\rm{dd}}/3\right)\chi_0(\mathbf{0})>0,\label{stablecond}
\end{equation}
where 
%\begin{equation}
$\chi_0(\mathbf{0})=\beta \zeta_{1/2}^+(e^{\beta[\mu-V_{\textrm{eff}}(\mathbf{0})]})/\ldb^3, $ %\label{chi00}
%\end{equation} 
(e.g.~see \cite{Mueller2000a}). It is worth noting that the $k\to0$ limit of the density response function is proportional to the compressibility, i.e.~$\kappa\!=\!n^{-2}\chi(\mathbf{k}\!\to\!0)$. Thus instability is signaled by diverging compressibility and hence density fluctuations of the system.

\begin{figure}[!tbh]
\begin{center}
    \includegraphics[width=3.4in]{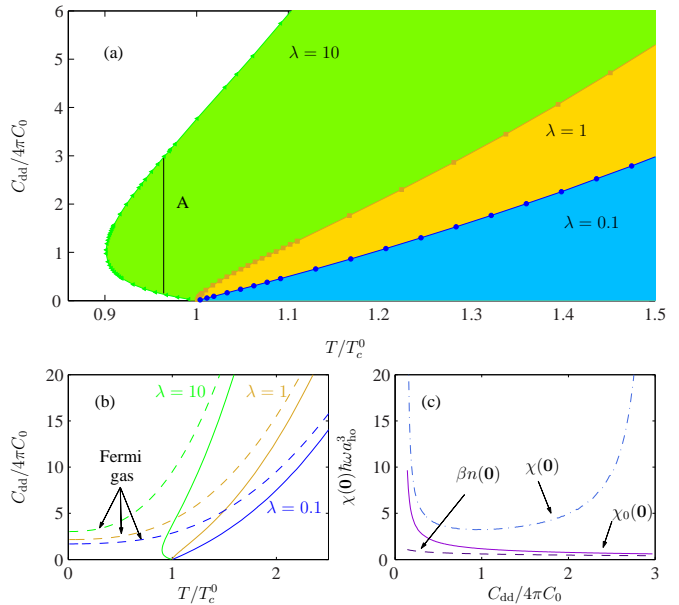}
\caption{(Color) Stability of a purely dipolar Bose gas. (a) Stability boundaries for different trap geometries, where the shaded regions to the  right of the boundaries are stable. Filled symbols indicate   self-consistent calculations of the stability boundary according to Eq.~(\ref{stablecond}). (b) Comparison of Bose (solid lines) and  Fermi (dashed lines) gas stability. (c) Density response functions for   $\lambda=10$   along the fixed temperature path A indicated in (a). The classical limit of the bare response function, $\beta n(\mathbf{0})$ is shown, where $n(\mathbf{0})$ is the density at $\mathbf{x}=\mathbf{0}$.
To make our calculations independent of $N$ we  scale $T$ by the ideal gas critical temperature $T_c^0 =\sqrt[3]{N/\zeta(3)}\hbar\omega/k_B$  and use the interaction parameter $C_0= {\hbar\omega a_{\rm{ho}}^3}/\sqrt[6]{N}$, where $\zeta(\alpha)\equiv\zeta_\alpha^+(1)$, $a_{\rm{ho}}=\sqrt{\hbar/m\omega}$ and $\omega=\sqrt[3]{\omega_{\rho}^2\omega_z}$.
 \label{Fig:StabGeom}}
\end{center}\vspace{-0.8cm}
\end{figure}

{\bf Results:} 
We present results for the stability regions of a purely dipolar gas ($g=0$) in Fig.~\ref{Fig:StabGeom}(a). 
We observe that the stability region grows with increasing $\lambda$.
The strong geometry dependence of these results arises from the anisotropy of the dipole interaction: In oblate geometries ($\lambda>1$) the dipoles are predominantly side-by-side and interact repulsively (stabilizing), whereas in prolate geometries ($\lambda<1$) the attractive (destabilizing) head-to-tail interaction of the dipoles dominates.
Geometry dependence has also been observed in the stability of a ($T=0$) dipolar Bose condensate \cite{Ronen2007a}.

A prominent feature in Fig.~\ref{Fig:StabGeom}(a) is that all the stability boundaries terminate  at the critical point with $\Cdd=0$. 
This can be understood because $\chi_0(\mathbf{0})$ of a saturated Bose gas diverges %[Eq.~(\ref{chi00})] 
and the system is unstable to the attractive component of the dipole interaction [see Eq.~(\ref{stablecond})].
The harmonic trap provides a long wavelength cut-off that limits the divergence of $\chi_0(\mathbf{0})$ \cite{Mueller2000a} and (beyond the semiclassical approximation) will allow systems with small $\Cdd$ to be stable below the critical point.
At temperatures well-above condensation  thermal pressure dominates and the critical dipole strength  for stability  increases with temperature.

It is interesting to contrast the behavior to that of an equivalent system with Fermi statistics, for which $n(\x)=\ldb^{-3}\zeta_{3/2}^-\left(e^{\beta[\mu-V_{\rm{eff}}(\x)]}\right)$. The stability regions, identified using an analogous procedure to the Bose gas, are shown in Fig.~\ref{Fig:StabGeom}(b). 
At high temperatures both systems exhibit similar stability properties.
The systems are distinctly different at low temperatures (note $T_F^0\approx1.93T_c^0$) with the Fermi system being stabilized by degeneracy pressure. We note that Fermi stability calculations using the same theory have been carried out in \cite{He2008a} for $T=0$.
 
A striking feature of the oblate system in Fig.~\ref{Fig:StabGeom}(a) is that the stability boundary bends back on itself so that the system is only stable for moderate values of $\Cdd$. This feature, which we refer to as \textit{double instability}, first arises for moderate anisotropies of the confining potential ($\lambda\gtrsim2$), but becomes more prominent as $\lambda$ increases.  
The physical origin of the double instability can be understood by considering system properties along the vertical line marked A in Fig.~\ref{Fig:StabGeom}(a), and by noting that in the purely dipolar case the stability condition (\ref{stablecond}) reduces to $\Cdd\chi_0(\mathbf{0})<3$.
For the lowest values of $\Cdd$ the system is saturated ($n\approx2.612/\ldb^3$ and $\chi_0\to\infty$) and unstable for any attractive interaction. 
However, since the average effect of the  dipolar interaction across the whole cloud is repulsive for the oblate trap, the effect of increasing
 $\Cdd$ is to decrease the central density. Any decrease in density from saturation causes a rather large decrease in $\chi_0$ [see  Fig.~\ref{Fig:StabGeom}(c)] and the system becomes quite stable (i.e.~$\chi$ takes a moderate value).
%Lowering the density (at constant $T$)  moves the system away from saturation, hence reducing $\chi_0$ and allowing the system to become more stable.
As $\Cdd$ is further increased the system eventually becomes unstable due to interactions  (noting  $\chi_0$ is rather small at this upper instability as the system is far from saturation)  [Fig.~\ref{Fig:StabGeom}(c)].
%The  density response functions and the peak density along path A is shown in Fig.~\ref{Fig:StabGeom}(c), 
Thus the lower instability arises from saturation (divergence of the bare response function) while the upper instability is driven by the large interaction strength. Because the normal Fermi gas cannot saturate, the double instability feature cannot occur for this system [c.f.~Fig.~\ref{Fig:StabGeom}(b)]

\begin{figure}[!tbh]
\begin{center}
%load cutoff;runVariational;plotAlphaBetaDt
    \includegraphics[width=3.2in]{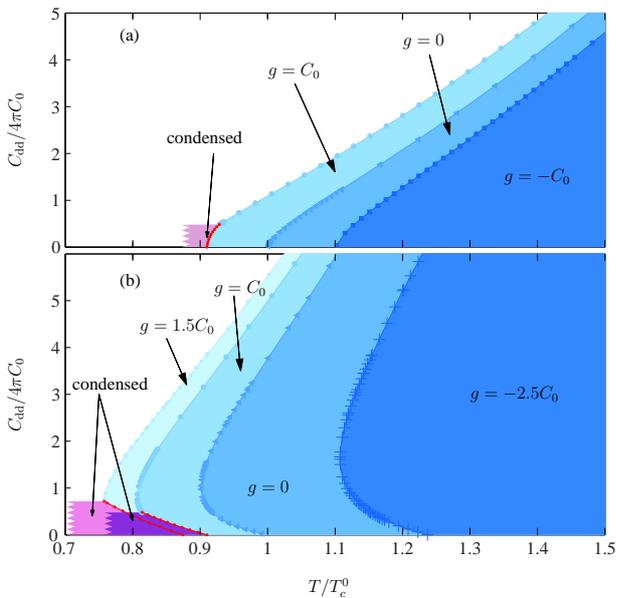}
\caption{(Color)  Stability for various values of the contact interaction with (a) $\lambda = 1$ and (b) $\lambda = 10$. Shaded regions to the right of the boundaries are stable. Symbols indicate where self-consistent calculations determine the stability boundary according to Eq.~(\ref{stablecond}). Stable condensed regions are indicated for cases with $g>0$ (for $g=0$ condensation occurs at $\Cdd=0$ with $T_c=T_c^0$, and for $g<0$ the system is unstable prior to condensation). Other parameters as  in Fig.~\ref{Fig:StabGeom}. 
 \label{Fig:StabC}}
\end{center}\vspace{-0.4cm}
\end{figure}

%\noindent{ \bf Contact Interactions:}\\
%\\
In Fig.~\ref{Fig:StabC} we investigate the effect of contact interactions on the stability diagram for two different trap geometries. In both cases we note that for repulsive contact interactions ($g>0$) the stability region is increased over the purely dipolar gas, while for the attractive case ($g<0$) stability is reduced. These observations can be qualitatively understood from the stability condition (\ref{stablecond}), e.g.~a positive value of $g$ can offset the attractive component of the DDI. Indeed, if $2g>\Cdd/3$ then there is no attractive component to the overall interaction and the system is stable. 
We see this in Fig.~\ref{Fig:StabC}(a) and (b) where the boundary lines terminate at the critical temperature $T_c$  with the dipole interaction strength $\Cdd^*=6g$. 
 In Fig.~\ref{Fig:StabC} we schematically indicate the stability regions for the condensed phase below $T_c$ using this result  \footnote{When the condensate fraction becomes appreciable we need to include the condensate density response, which gives the low temperature stability condition  $\Cdd^*=3g$.  This is the Thomas-Fermi stability condition  \cite{Eberlein2005a} which is the same as for the homogeneous gas and does not include quantum pressure  \cite{Ronen2007a}.}.
 
Finally we comment on the effect of contact interactions on the double instability feature that occurs in the oblate trap [Fig.~\ref{Fig:StabC}(b)]. Attractive contact interactions make the feature more prominent (noting  this case cannot stably condense). The double instability region gets smaller for moderate values of repulsive contact interactions and admits a stable condensate phase. For sufficiently large values of $g$ the double instability region disappears. In the $\lambda=10$ trap this occurs at $g\approx1.5C_0$ [Fig.~\ref{Fig:StabC}(b)].
 
{\bf Discussion: }  
Our results are most significantly applicable to current experiments with polar molecular gases, which are now approaching degeneracy.
%While progress has been most rapid with fermionic $^{40}$K$^{87}$Rb \cite{Ni2008a}, recently ground state bosonic molecules of $^{41}$K$^{87}$Rb have been produced \cite{Aikawa2010a}. 
To put our results in context we now discuss the typical interaction parameters accessible in the lab. While the dipole strength is dependent on the electric field applied, using the maximal value for KRb of $d\approx 0.57$ D we have $\Cdd\approx 4.76\times4\pi C_0$ \footnote{We have taken $N=10^5$ and $\omega=2\pi\times100$ Hz.}. For RbCs the dipole moment is expected to be about twice that of KRb \cite{Kotochigova2005a}.
Much less is understood about the contact interactions of the molecular systems, although some progress on understanding the s-wave properties has been made in Ref.~\cite{Ospelkaus2010a}.

In contrast, atomic systems  interact with typically much weaker magnetic dipoles. For comparison (using the same trap and particle number) the parameters of $^{52}$Cr would be $\Cdd= 0.0117 \times4\pi C_0 $  %(0.0238)
and $g=0.325 C_0$ %(0.658)
 (which can be varied using a Feshbach resonance \cite{Koch2008a}). 
 This value of $\Cdd$ is rather small on the scale of our phase diagrams [Figs.~\ref{Fig:StabGeom}(a)  and \ref{Fig:StabC}(a)-(b)], and thus instability above $T_c$ is not a concern for this system even in the absence of any contact interaction. Furthermore,  since $g\gg \Cdd/6$ this system is stable below $T_c$.
 As noted earlier, for the trapped system the bare compressibility gets large but does not diverge at $T_c$ as the residual quantum pressure provides some  stability.
An investigation of condensate stability was performed using the Gross-Pitaevskii equation \cite{Ronen2007a}.
There it was found that stability depends on $\lambda$ in a complicated manner but mostly the stability boundary was seen to increase with increasing $\lambda$, reaching only $\Cdd \approx 0.006\times 4\pi C_0$ for $\lambda = 10$ and $N=10^5$.
 This also allows us to conclude that while quantum pressure is an important consideration for the small atomic dipoles it is rather unimportant for the above polar molecules at their maximal dipole strength.

{\bf Conclusions:} 
In this paper we have quantified the  stability of a normal  dipolar Bose gas. Our results quantify the rich interplay of DDI anisotropy, trap geometry and contact interactions in determining the stability regions, and demonstrate the distinctive behavior of the  dipolar  Bose and Fermi gases. 
 We have predicted  a novel double instability region  in oblate traps, and  explained how this arises from a competition between saturation and interaction effects. 
 
 Pivotal to our analysis has been the use of the density response function in the RPA, which we find to accurately predict where our meanfield calculations become unstable and provide a quantitative condition for stability of the saturated gas. Interestingly, experimental techniques have recently been developed to measure the  density response function (equivalently the compressibility) in a trapped atomic gas \cite{Sanner2011a,*Hung2011a,*Tung2010a}. Using these techniques should furnish a broader understanding of the mechanisms leading to instability.

Our predictions will be relevant to current and emerging experiments with polar molecules and suggest that a range of strategies including the use of highly oblate traps, reduction of the dipole strength, or increasing the contact interaction strength will be necessary to have a stable pathway to cool the system to a Bose-Einstein condensate. Additionally, our prediction of a novel double instability feature occurs in oblate traps currently favored in experiments (e.g.~see \cite{Miranda2011a}).

There are several extensions to the theory  that will be considered in future work: (i) Beyond semiclassical treatment of modes to account for the quantum pressure effects below $T_c$; (ii) The inclusion of dipolar exchange interactions; (iii) Beyond meanfield treatment to account for correlation in the strong dipole regime.
Classical field methods provide an avenue of investigation that accounts for many of these considerations and should be applicable  at and below $T_c$  \cite{Blakie2009e}.

{\bf Acknowledgments:}  This work was supported by Marsden contract UOO0924, and FRST contract NERF-UOOX0703.

\bibliographystyle{apsrev4-1}

%\bibliography{dipolar}

%Merlin.mbs v4.21 2009-07-09.
%

\end{document}